\documentclass[aps,draft,twocolumn]{revtex4}

\usepackage{color}
\input epsf

\begin{document}

\title{
One-Dimensional Fermions with neither Luttinger-Liquid nor Fermi-Liquid
Behavior 
}
\author{A.V. Rozhkov}

\affiliation{
Institute for Theoretical and Applied Electrodynamics of Russian Academy of
Science, ul.~Izhorskaya~13, Moscow, 125412, Russia
}

\affiliation{
Moscow Institute of Physics and Technology, Institutskiy per. 9,
Dolgoprudny, Moscow Region, 141700, Russia
}

\date{\today}

%\maketitle

\begin{abstract}
It is well-known that, generically, the one-dimensional interacting
fermions cannot be described in terms of the Fermi liquid. Instead, they
present different phenomenology, that of the Tomonaga-Luttinger liquid: the
Landau quasiparticles are ill-defined, and the fermion occupation number is
continuous at the Fermi energy. We demonstrate that suitable fine-tuning of
the interaction between fermions can stabilize a peculiar state of
one-dimensional matter, which is dissimilar to both the Tomonaga-Luttinger
and Fermi liquids. We propose to call this state a quasi-Fermi liquid.
Technically speaking, such liquid exists only when the fermion interaction
is irrelevant (in the renormalization group sense). The quasi-Fermi liquid
exhibits the properties of both the Tomonaga-Luttinger liquid and the Fermi
liquid. Similar to the Tomonaga-Luttinger liquid, no finite-momentum
quasiparticles are supported by the quasi-Fermi liquid; on the other hand,
its fermion occupation number demonstrates finite discontinuity at the
Fermi energy, which is a hallmark feature of the Fermi liquid. Possible
realization of the quasi-Fermi liquid with the help of cold atoms in an
optical trap is discussed.
\end{abstract}

\maketitle
%\hfill
\textit{Introduction.}-- An important goal of the modern many-body physics
is the search for exotic states of matter. Appropriate examples are spin
liquids
\cite{spin_liq1,spin_liq2,spin_liq3},
Majorana fermion
\cite{majorana1,majorana2,majorana3,majorana4,majorana5},
topological insulators and semimetals
\cite{top_ins1,top_ins2,weyl1},
and others. A peculiar state of one-dimensional (1D) fermionic matter
deviating from known types of interacting Fermi systems is the subject of
this paper.

Let us remind ourselves that the most basic model of the interacting
fermions is that of the Fermi liquid. It successfully describes a variety
of interacting fermion systems (e.~g., electrons in solids, atoms
of helium-3)
\cite{agd}. 
The approach is based on the Landau's conjecture that both the ground state
of a Fermi liquid and its low-lying excitations are adiabatically connected
to states of the non-interacting Fermi gas. If the interaction is weak,
this hypothesis implies that the perturbation theory in the interaction
strength is valid. The latter supplies a theorist with a tool to study
specific examples.

A known system for which the Landau conjecture fails is a 1D liquid of
interacting fermions. The interacting 1D fermions constitute
a separate universality class, so-called Tomonaga-Luttinger liquid
\cite{boson,book_giamarchi}:
unlike the Fermi liquid, the Tomonaga-Luttinger ground and excited states
have zero overlap with the corresponding non-interacting states, the
Tomonaga-Luttinger liquid properties cannot be calculated perturbatively
with interaction strength as a small parameter.

In 1D the Tomonaga-Luttinger liquid is a generic state of matter.
However, recent progress in fabrication and control over the properties of
the many-particle systems allows us to look for more fragile types of 1D
correlated liquids. Specifically, consider a gas of Fermi atoms in a 1D
trap 
\cite{1d_fermi_atoms_2005}. 
It is within modern experimental capabilities to vary the effective
interaction constant of the optically trapped atoms, and even tune the
constant to zero
\cite{cold_atoms_rmp2008,ultracold_fermi_rmp2008}.
Below we will demonstrate that such nullification of the {\it effective}
coupling constant does not imply vanishment of all {\it microscopic}
interactions. Some residual interactions remain, and in 1D they
stabilize a peculiar state of matter, which we propose to call a
quasi-Fermi liquid. The latter state appears to be a hybrid of both the
Fermi and the Tomonaga-Luttinger liquids: its ground state is
perturbatively connected to the ground state for the free fermions, yet,
the perturbatively-defined quasiparticles do not exist. That is, in case of
the quasi-Fermi liquid the Landau conjecture valid only for the ground
state, but not for excitations. Of course, there is nothing special about
the cold atoms, and the quasi-Fermi liquid may be realized in other fermion
systems, which allow adequate fine-tuning of the coupling.

%In this paper we discuss, using perturbation theory, the condition
%necessary for the quasi-Fermi liquid existence, explain its relation to the
%generic Tomonaga-Luttinger liquid, and study the ground state and
%excitations of the quasi-Fermi liquid.

The presentation below has the following structure. First, we formally
introduce our model. Second, the self-energy is evaluated perturbatively,
which allows to determine both the quasiparticle residue and the occupation
number corrections. Third, analyzing these quantities we will be able to
define the quasi-Fermi liquid as a distinct state of fermionic matter.
Fourth, we discuss possible implementation of such quantum liquid using
optically trapped cold atoms. Finally, we formulate our conclusions. In the
Supplemental Material we present the extension of our calculations
beyond the second-order perturbation theory, and discuss other subtleties.

\textit{The studied model.}--
The 1D interacting fermions are commonly described by the
Tomonaga-Luttinger Hamiltonian:
\begin{eqnarray}
H_{\rm TL}=H_{\rm kin} + H_{\rm int},
%%%%%%%%%%%%%%%%%%%%%%%%%%%%%%%%%%%%%%%%%%%%%%%%%%
\label{H_TL}
%%%%%%%%%%%%%%%%%%%%%%%%%%%%%%%%%%%%%%%%%%%%%%%%%% 
\\
H_{\rm kin} = i v_{\rm F}\int dx
\left( \colon\psi^\dagger_{{\rm L}}
\nabla\psi^{\vphantom{\dagger}}_{{\rm L}}\colon - \colon\psi^\dagger_{{\rm
R}}
\nabla\psi^{\vphantom{\dagger}}_{{\rm R}}\colon \right) ,\label{H_kin}\\
H_{\rm int} = g \int dx \rho_{\rm L} \rho_{\rm R},
%%%%%%%%%%%%%%%%%%%%%%%%%%%%%%%%%%%%%%%%%%%%%%%%%% 
\label{int}
%%%%%%%%%%%%%%%%%%%%%%%%%%%%%%%%%%%%%%%%%%%%%%%%%% 
\end{eqnarray}
where
$\psi_p$,
is the field operator for the right-moving ($p$=R) and left-moving ($p$=L)
fermions, operators
$\rho_p = \colon
			\psi_p^\dagger 
			\psi_p^{\vphantom{\dagger}}
		\colon
$
are the densities of the left- and right-movers,
$v_{\rm F}$ is the Fermi velocity, $g$ is the coupling constant. Colons
denote the normal ordering.

The Tomonaga-Luttinger liquid differs from the Fermi liquid: the
perturbatively-defined quasiparticles are absent, the Fermi occupation
number 
$n^p_{k} = \langle c^\dag_{pk} c^{\vphantom{\dagger}}_{pk} \rangle$
has no discontinuity at the Fermi point, the Tomonaga-Luttinger ground
state has zero overlap with the free fermion ground state.

The culprit responsible for these abnormalities is the fermion-fermion
interaction
$
H_{\rm int}
$,
which is marginal in the renormalization group sense. The perturbation
theory in orders of $g$ has additional divergences absent in the
higher-dimensional systems. For example, the Matsubara single-particle
self-energy is equal to 
\cite{shitov_levitov,rozhkov_spinless_variational,xover_2012}
\begin{eqnarray}
%%%%%%%%%%%%%%%%%%%%%%%%%%%%%%%%%%%%%%%%%%%%%%%%%%
\label{lutt_self-energy}
%%%%%%%%%%%%%%%%%%%%%%%%%%%%%%%%%%%%%%%%%%%%%%%%%% 
\Sigma^p_{\rm TL}
&=&
\frac{g^2}{16 \pi^2 v_{\rm F}^2}
(i\nu - p v_{\rm F} k )
%\\
%\nonumber
%&&\times
\ln \left(
                \frac{
                        v_{\rm F}^2 k^2 + \nu^2
                     }
                     {
                        4 v_{\rm F}^2 \Lambda^2
                     }
    \right) + \ldots,
\end{eqnarray}
where the ellipsis stands for the less singular terms, 
$p = +1$ 
($p = -1$)
for the right-moving (left-moving) fermions, and $\Lambda$ is the
ultraviolet cutoff. This self-energy corresponds to the following
expression 
\begin{eqnarray} 
%%%%%%%%%%%%%%%%%%%%%%%%%%%%%%%%%%%%%%%%%%%%%%%%%%
\label{Z_1d}
%%%%%%%%%%%%%%%%%%%%%%%%%%%%%%%%%%%%%%%%%%%%%%%%%% 
\delta Z^p_{\rm TL}
=
\frac{g^2}{16 \pi^2 v_{\rm F}^2}
\ln \left(
                \frac{
                        4 v_{\rm F}^2\Lambda^2
                }
                {
                        v_{\rm F}^2 k^2 + \nu^2
                }
    \right) + \ldots,
\end{eqnarray}
for correction to the quasiparticle residue 
$Z^p_{\rm TL} = 1 - \delta Z^p_{\rm TL}$.
The correction diverges for small $\nu$ and $k$. As a result, the
conventional Fermi
quasiparticles are ill-defined, and the occupation number function has a
power-law singularity instead of the discontinuity. The properties of
$H_{\rm TL}$,
Eq.~(\ref{H_TL}),
are now well-understood
\cite{boson,book_giamarchi}.

However, it is sometimes required to include the irrelevant operators into
consideration. There are two least irrelevant operators:
\begin{eqnarray}
H_{\rm nl} 
= 
v'_{\rm F}
\int dx \left[
		\colon\! 
			(\nabla\psi^\dagger_{\rm L})
			(\nabla \psi_{\rm L}^{\vphantom{\dagger}})
		\colon 
		+
		\colon\! 
			(\nabla\psi^\dagger_{\rm R})
			(\nabla \psi_{\rm R}^{\vphantom{\dagger}})
		\colon 
        \right],  
%%%%%%%%%%%%%%%%%%%%%%%%%%%%%%%%%%%%%%%%%%%%%%%%%%
\label{H2}
%%%%%%%%%%%%%%%%%%%%%%%%%%%%%%%%%%%%%%%%%%%%%%%%%% 
\\
H_{\rm int}' = ig' \int dx 
 \left\{
	\rho_{\rm R} 
	\left[ 
		\colon 
			\psi^\dagger_{\rm L}
			(\nabla \psi^{\vphantom{\dagger}}_{\rm L}) 
		\colon 
		-
		\colon 
			(\nabla \psi^\dagger_{\rm L}) 
			\psi^{\vphantom{\dagger}}_{\rm L}
		\colon 
	\right]
\right.
\\
%%%%%%%%%%%%%%%%%%%%%%%%%%%%%%%%%%%%%%%%%%%%%%%%%%
\label{H'}
%%%%%%%%%%%%%%%%%%%%%%%%%%%%%%%%%%%%%%%%%%%%%%%%%% 
\left. 
-  
	\rho_{\rm L} 
	\left[ 
		\colon 
			\psi^\dagger_{\rm R}
			(\nabla \psi^{\vphantom{\dagger}}_{\rm R}) 
		\colon
		-
		\colon 
			(\nabla \psi^\dagger_{\rm R}) 
			\psi^{\vphantom{\dagger}}_{\rm R}
		\colon 
	\right]
\right\}.
\nonumber 
\end{eqnarray}
Here
$H_{\rm nl}$
is the quadratic correction to the linear dispersion of the fermions,
$H_{\rm int}'$
is the irrelevant interaction. Both 
$H_{\rm nl}$
and
$H_{\rm int}'$
have the scaling dimension of 3 (the dimension of the gradient operator is
1, each field operator has the dimension of 1/2). Other irrelevant
operators have higher scaling dimensions, therefore, their effects are less
pronounced.

Recently, the Hamiltonian
\begin{eqnarray} 
H = H_{\rm TL} + H_{\rm nl} + H_{\rm int}'
%%%%%%%%%%%%%%%%%%%%%%%%%%%%%%%%%%%%%%%%%%%%%%%%%%
\label{H_gen}
%%%%%%%%%%%%%%%%%%%%%%%%%%%%%%%%%%%%%%%%%%%%%%%%%%
\end{eqnarray} 
and its modifications have been investigated actively
\cite{samokhin1998,pustilnik_etal2003,rozhkov2005,pereira_etal2006,
rozhkov2006,pustilnik_etal2006,teber2006,teber2007,pereira_etal2007,
pirooznia_kopietz2007,khodas_etal2007,aristov2007,rozhkov_prb2008,
pirooznia_etal2008,imambekov_glazman2009,imambekov_glazman_science2009,
pereira_sela2010,ivanov2010,schmidt_etal2010}.
These studies have demonstrated that combined effect of the marginal and
the irrelevant operators has important and measurable consequences for
system's properties.

In this paper we will discuss the model of the 1D fermions without the
marginal interaction at all:
\begin{eqnarray} 
H_{\rm ii} = H_{\rm kin} + H_{\rm nl} + H_{\rm int}',
%%%%%%%%%%%%%%%%%%%%%%%%%%%%%%%%%%%%%%%%%%%%%%%%%%
\label{no_marg}
%%%%%%%%%%%%%%%%%%%%%%%%%%%%%%%%%%%%%%%%%%%%%%%%%% 
\end{eqnarray} 
where `ii' stands for `irrelevant interaction'. We may name two examples
where
$H_{\rm ii}$
is applicable. First, consider the cold Fermi atoms in a 1D trap. Under
rather general conditions the suitable Hamiltonian is given by 
Eq.~(\ref{H_TL}),
see
Refs.~\cite{ultracold_fermi_rmp2008,olshanii1998}.
However, the interaction between the atoms is highly adjustable
\cite{ultracold_fermi_rmp2008,cold_atoms_rmp2008},
which may be used to our advantage: below we will offer an argument
suggesting that the system parameters can be tuned in such a manner that
$g$ [or, more precisely, renormalized coupling 
$g^{\rm eff} = g + {\cal O} ((g')^2)$] 
vanishes, but
$g' \ne 0$.

Our second case requires no fine-tuning. Using the unitary transformation
of
Ref.~\onlinecite{mattias_lieb1965},
it has been demonstrated that the Tomonaga-Luttinger Hamiltonian with
non-linear dispersion,
Eq.~(\ref{H_gen}),
may be
mapped~\cite{rozhkov2005,rozhkov2006,rozhkov_prb2008}
on Hamiltonian 
$H_{\rm ii}$
(see also
Ref.~\onlinecite{explanation_mapping}).
Therefore, the properties of 
$H_{\rm ii}$
are important for the theoretical description of the generic model $H$.

Superficially, one expects that, since 
$H_{\rm ii}$
has only the irrelevant interaction, it describes a kind of 1D Fermi
liquid. Indeed, using the perturbation theory, we will demonstrate that the
correction to the fermion occupation number
$n_k^p$
is finite and small. However, in a drastic departure from the Fermi liquid
picture, the quasiparticle residue correction diverges on the mass surface.
Thus, Hamiltonian
$H_{\rm ii}$
describes a state of 1D matter which lies halfway between the Fermi liquid
and the Tomonaga-Luttinger liquid: 
$n_k^p$
has the finite discontinuity at the Fermi energy, but no perturbatively
defined quasiparticles exist. This is our quasi-Fermi liquid.

%The paper is organized as follows. In
%Sec.~\ref{pert_theor}
%we derive the perturbative corrections to the quasiparticle residue and to
%the fermion occupation function. In
%Sec.~\ref{qfl}
%we discuss these findings and define the quasi-Fermi liquid. Matters
%related to the experimental implementation and identification of the
%quasi-Fermi liquid are discussed in
%Sec.~\ref{experiment}.
%In Sec.~\ref{conclusions}
%the conclusions are given.
%--------------------------------------------------------------
\begin{figure} [t]
\centering
\leavevmode
\epsfysize=5cm
\epsfbox{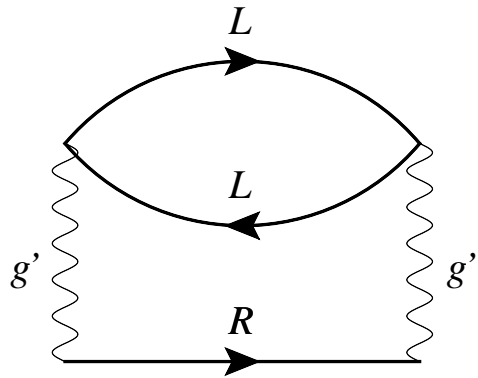}
\caption[]
{\label{fig::diag} 
The leading self-energy correction diagram. The solid lines with arrows
and `L', `R' chirality labels correspond to the fermion propagators. The
wiggly lines are irrelevant interactions.
}
\end{figure}
%--------------------------------------------------------------

\textit{Self-energy correction.}--
%%%%%%%%%%%%%%%%%%%%%%%%%%%%%%%%%%%%%%%%%%%%%%%%%%
%\label{pert_theor}
%%%%%%%%%%%%%%%%%%%%%%%%%%%%%%%%%%%%%%%%%%%%%%%%%% 
To implement the outlined plan, we must calculate the self-energy. For
definiteness, consider the self-energy for right-movers. Corresponding
diagram is shown in 
Fig.~\ref{fig::diag}.

The expression which must be evaluated is
\begin{eqnarray}
%%%%%%%%%%%%%%%%%%%%%%%%%%%%%%%%%%%%%%%%%%%%%%%%%%
\label{self-energy}
%%%%%%%%%%%%%%%%%%%%%%%%%%%%%%%%%%%%%%%%%%%%%%%%%% 
\Sigma^{\rm R}_{k, i\nu}
=
-
(g')^2 T^2 \sum_{i \Omega, i\nu'}
\int_{Q,q} (2q - 2k )^2
G^{\rm R, 0}_{k-Q, i\nu - i \Omega}
\\
\nonumber  
\times
G^{\rm L, 0}_{q-Q, i \nu'}
G^{\rm L, 0}_{q, i \Omega + i \nu'}.
\end{eqnarray} 
In this equation
$\int_{k} \ldots = \int (dk/2\pi) \ldots$;
the free Matsubara propagator is
$
G^{p, 0}_{k, i \omega}
=
( i \omega - \varepsilon_{k}^p )^{-1},
$
where the fermion dispersion is 
%\begin{eqnarray} 
$\varepsilon_{k}^p = p v_{\rm F} k + v_{\rm F}'k^2$.
%\end{eqnarray} 
The factor
$(2k - 2q)^2$
appears because each interaction line contributes a factor of
$g'(2k-2q)$
to the diagram. The overall minus sign accounts for the presence of a
single fermion loop. Calculating momentum integrals we assume that
\begin{eqnarray}
%%%%%%%%%%%%%%%%%%%%%%%%%%%%%%%%%%%%%%%%%%%%%%%%%%
\label{interaction_cutoff}
%%%%%%%%%%%%%%%%%%%%%%%%%%%%%%%%%%%%%%%%%%%%%%%%%% 
|q|, |Q| < \Lambda
<k_{\rm F}
=
\frac{v_{\rm F}}{2v_{\rm F}'},
\end{eqnarray} 
where
$k_{\rm F}$
is the Fermi momentum. This way we may avoid complications arising from
spurious zeros of
$\varepsilon^{p}_{k}$
which are located at
$k=-2 p k_{\rm F}$.
%(These zeros must be disregarded since our formula for
%$\varepsilon^{p}_{k}$
%is invalid beyond small-momentum range.) 

Performing the standard summation over $i \Omega$ and $i \nu$ and
taking the limit
$T \rightarrow 0$
we find
%\begin{widetext}
\begin{eqnarray}
%%%%%%%%%%%%%%%%%%%%%%%%%%%%%%%%%%%%%%%%%%%%%%%%%%
\label{self-energy2}
%%%%%%%%%%%%%%%%%%%%%%%%%%%%%%%%%%%%%%%%%%%%%%%%%% 
\Sigma^{\rm R}
=
(g')^2
\int_{Q,q}
(2q - 2k)^2 
\left[ 
	\theta(-\varepsilon^{\rm L}_{q}) 
	-
	\theta(-\varepsilon^{\rm L}_{q-Q}) 
\right]
\\
\nonumber
\times
\frac{
	\theta
	(
		\varepsilon^{\rm L}_{q-Q}
		-
		\varepsilon^{\rm L}_{q}
	)
	-
	\theta
	(
		\varepsilon^{\rm R}_{k-Q}
	)
}
{
	i\nu
	-
	\varepsilon^{\rm R}_{k-Q}
	-
	\varepsilon^{\rm L}_{q}
	+
	\varepsilon^{\rm L}_{q-Q}
}.
\end{eqnarray} 
For our purposes it is convenient to evaluate the imaginary part of the
retarded self-energy:
\begin{eqnarray}
%%%%%%%%%%%%%%%%%%%%%%%%%%%%%%%%%%%%%%%%%%%%%%%%%%
\label{Im_self-energy}
%%%%%%%%%%%%%%%%%%%%%%%%%%%%%%%%%%%%%%%%%%%%%%%%%% 
{\rm Im \,} \Sigma^{\rm R}_{\rm ret}
=
-\pi (g')^2
\int_{Q,q}
(2q - 2k)^2 
\left[ 
	\theta(-\varepsilon^{\rm L}_{q}) 
	-
	\theta(-\varepsilon^{\rm L}_{q-Q}) 
\right]
\nonumber 
\\
\nonumber
\times
\left[
	\theta
	(
		\varepsilon^{\rm L}_{q-Q}
		-
		\varepsilon^{\rm L}_{q}
	)
	-
	\theta
	(
		\varepsilon^{\rm R}_{k-Q}
	)
\right]
\delta 
\left(
	\nu
	-
	\varepsilon^{\rm R}_{k-Q}
	-
	\varepsilon^{\rm L}_{q}
	+
	\varepsilon^{\rm L}_{q-Q}
\right).
\\
\end{eqnarray} 
%\end{widetext}
Now we integrate over $Q$:
%\begin{widetext}
\begin{eqnarray}
%%%%%%%%%%%%%%%%%%%%%%%%%%%%%%%%%%%%%%%%%%%%%%%%%%
\label{Im_self-energy2}
%%%%%%%%%%%%%%%%%%%%%%%%%%%%%%%%%%%%%%%%%%%%%%%%%% 
{\rm Im \,} \Sigma^{\rm R}_{\rm ret}
=
-(g')^2
\int_{q}
\frac
{
	(k - q)^2 
}
{
	v_{\rm F} + v_{\rm F}'(k - q)
}
%\qquad
%\\
\nonumber
%\times
\left[ 
	\theta(q) 
	-
	\theta\left(q-Q^*\right) 
\right]
\\
%\nonumber
\times
\left[
	\theta
	(
		\varepsilon^{\rm L}_{q-Q^*}
		-
		\varepsilon^{\rm L}_{q}
	)
	-
	\theta
	(
		\varepsilon^{\rm R}_{k-Q^*}
	)
\right],
\qquad
%\\
%\nonumber
%\times
\end{eqnarray} 
%\end{widetext}
where
$Q^* (q)$
delivers zero to the argument of the delta-function in
Eq.~(\ref{Im_self-energy}):
\begin{eqnarray} 
%%%%%%%%%%%%%%%%%%%%%%%%%%%%%%%%%%%%%%%%%%%%%%%%%%
\label{root}
%%%%%%%%%%%%%%%%%%%%%%%%%%%%%%%%%%%%%%%%%%%%%%%%%% 
Q^*
=
-
\frac{
	\Delta \nu
     }
     {
	2 v_{\rm F} + 2 v_{\rm F}'(k-q)
     },
\qquad
\Delta \nu = \nu - \varepsilon^{\rm R}_k.
\end{eqnarray}
Thus, we need to evaluate the integral
\begin{eqnarray}
%%%%%%%%%%%%%%%%%%%%%%%%%%%%%%%%%%%%%%%%%%%%%%%%%%
\label{integral}
%%%%%%%%%%%%%%%%%%%%%%%%%%%%%%%%%%%%%%%%%%%%%%%%%% 
I = 
\int_{0}^{q^*} dq
\frac
{
	(k - q)^2 
	\left[
		\theta
		(
			\varepsilon^{\rm L}_{q-Q^*}
			-
			\varepsilon^{\rm L}_{q}
		)
		-
		\theta
		(
			{k - Q^*}
		)
	\right]
}
{
	v_{\rm F} + v_{\rm F}'(k - q)
},
\end{eqnarray} 
where upper limit of integration
%\begin{eqnarray} 
$q^*
\approx
-
\Delta \nu / 2 v_{\rm F} $
%\frac{ \Delta \nu }{ 2 v_{\rm F} }
%\end{eqnarray} 
satisfies the equation
$q^* = Q^*(q^*)$.
It is easy to check that
\begin{eqnarray}
\theta
(
	\varepsilon^{\rm L}_{q-Q^*}
	-
	\varepsilon^{\rm L}_{q}
)
=
\theta
(
	Q^* [ v_{\rm F} - v_{\rm F}' (2q - Q^*) ]
)
=
\theta (Q^*).
\end{eqnarray} 
Further, analyzing
Eq.~(\ref{root}),
we determine that the sign of 
$Q^*$
coincides with the sign of
$(\varepsilon_k^{\rm R} - \nu)$.
Consequently,
\begin{eqnarray}
\theta
(
	\varepsilon^{\rm L}_{q-Q^*}
	-
	\varepsilon^{\rm L}_{q}
)
=
\theta
(
	\varepsilon_k^{\rm R} - \nu
),
\end{eqnarray} 
where the function on the right-hand side is independent of the integration
variable $q$.

The second step-function
$
\theta ( { k - Q^*})
$
in 
Eq.~(\ref{integral}) 
can be evaluated easily near the mass surface
$\nu = \varepsilon_{k}^{\rm R}$.
When the mass surface is approached,
$Q^* \rightarrow 0$;
consequently,
$\theta ( k - Q^* ) = \theta(k)$.

Since both step-functions are independent of the integration variable $q$,
the integral $I$ can be trivially evaluated to the lowest order in
$\nu - \varepsilon_k^{\rm R}$.
Keeping the most singular term, we derive
\begin{eqnarray}
%%%%%%%%%%%%%%%%%%%%%%%%%%%%%%%%%%%%%%%%%%%%%%%%%%
\label{Im_Sigma}
%%%%%%%%%%%%%%%%%%%%%%%%%%%%%%%%%%%%%%%%%%%%%%%%%% 
{\rm Im\,} \Sigma_{\rm ret}^{\rm R}
=
-
\frac{(g' k)^2}{4\pi v_{\rm F}^2} 
(\varepsilon_k^{\rm R} - \nu)
\left[
	\theta
	(
		\varepsilon_k^{\rm R} - \nu
	)
-
	\theta (k)
\right]
+
\delta \Sigma, \quad
%\ldots, \quad
%\\
%\nonumber
%=
%-
%\frac{(g' k)^2}{4\pi v_{\rm F}^2} 
%|\nu - \varepsilon_k^{\rm R}|
%	\theta
%	(
%		k(\nu - \varepsilon_k^{\rm R})
%	)
%+
%\ldots
\end{eqnarray} 
where 
$\delta \Sigma$
stands for less singular terms.
%Deriving this formula we used the identity:
%\begin{eqnarray}
%a[\theta(a) - \theta(b)] = |a|\theta(-ab).
%\end{eqnarray} 
To obtain
${\rm Re\,} \Sigma_{\rm ret}^{\rm R}$
we use Kramers-Kronig relations. For the first term in
Eq.~(\ref{Sigma})
the Kramers-Kronig integral can be easily calculated analytically (with 
$\Lambda$ 
playing the role of the high-energy cutoff):
\begin{eqnarray}
%%%%%%%%%%%%%%%%%%%%%%%%%%%%%%%%%%%%%%%%%%%%%%%%%%
\label{Sigma}
%%%%%%%%%%%%%%%%%%%%%%%%%%%%%%%%%%%%%%%%%%%%%%%%%% 
\Sigma_{\rm ret}^{\rm R}
=
\frac{(g'k)^2}{4\pi^2 v_{\rm F}^2} 
(\nu - \varepsilon_k^{\rm R})
\ln
\left(
	\frac{
		\nu - \varepsilon_k^{\rm R} + i0
	     }
	     {
		v_{\rm F} \Lambda
	     }
\right)
+ \ldots
\end{eqnarray} 
The less-singular contribution due to 
$\delta \Sigma$ 
is replaced by the ellipsis.

Equation~(\ref{Sigma})
resembles 
Eq.~(\ref{lutt_self-energy}):
both have singularities at the mass surface. Yet, there is an important
difference: expression in
Eq.~(\ref{Sigma})
has an extra $k^2$ factor, which acts to weaken the singular contribution
at small $k$. We will see that peculiar properties of our system may be
traced back to this feature of the self-energy.

The quasiparticle residue
\cite{mahan} 
\begin{eqnarray}
Z^{\rm R} (k)
=
\left.
\frac{1}
     { 1 - \partial \Sigma^{\rm R}_{\rm ret}/ \partial \nu}
\right|_{\nu = \varepsilon_k^{\rm R}},
\\
%%%%%%%%%%%%%%%%%%%%%%%%%%%%%%%%%%%%%%%%%%%%%%%%%%
\label{dZ}
%%%%%%%%%%%%%%%%%%%%%%%%%%%%%%%%%%%%%%%%%%%%%%%%%% 
\frac{
	\partial \Sigma^{\rm R}_{\rm ret}
     }
     {
	\partial \nu
     }
=
\frac{(g'k)^2}{4\pi^2 v_{\rm F}^2} 
\ln
\left(
	\frac{
		\nu - \varepsilon_k^{\rm R} + i0
	     }
	     {
		v_{\rm F} \Lambda
	     }
\right)
+ \ldots
\end{eqnarray}
vanishes for any finite $k$ due to the divergence of 
$\partial \Sigma^{\rm R}_{\rm ret}/ \partial \nu$
on the mass surface. Thus, like the Tomonaga-Luttinger model, our system
does not support the perturbatively-defined quasiparticles. However, since
the interaction is irrelevant, in the Matsubara domain 
$\partial \Sigma^{\rm R}/ \partial \nu$
remains finite, while expression in
Eq.~(\ref{Z_1d})
diverges. The divergence is sensitive to temperature: if one replaces the
step-functions in
Eq.~(\ref{Im_self-energy})
by appropriate Fermi functions, the resultant 
${\rm Im}\, \Sigma_{\rm ret}^{\rm R}$
becomes continuous function of its arguments. Consequently, 
${\rm Re}\, \Sigma_{\rm ret}^{\rm R}$
becomes finite. Note also, that
Eq.~(\ref{dZ}) 
does not contain $v_{\rm F}'$ explicitly. Thus, the destruction of
the quasiparticles occurs even for systems with linear dispersion, as long
as interaction is non-zero
$g' \ne 0$.

Despite the absence of the quasiparticles, the fermionic occupation numbers
$n_k^p = \langle c^\dag_{pk} c^{\vphantom{\dagger}}_{pk} \rangle$
remain well-defined. This is not surprising: any finite-order correction
to a ground-state matrix element due to irrelevant interaction is finite
(since $Z_k$ is a property of an excited state, it is exempt from this
rule). To calculate
$\delta n_k^p$
explicitly we start with the formula
%\begin{eqnarray}
$n_k^{\rm R} = -\int_{- \infty}^{0} \frac{d \nu}{\pi}
{\rm Im\,} G^{\rm R}_{{\rm ret},k,\nu}$.
%\end{eqnarray}
Therefore, the second-order correction is equal to
\begin{eqnarray}
%%%%%%%%%%%%%%%%%%%%%%%%%%%%%%%%%%%%%%%%%%%%%%%%%%
\label{occup_corr}
%%%%%%%%%%%%%%%%%%%%%%%%%%%%%%%%%%%%%%%%%%%%%%%%%% 
\delta n_k^{\rm R} = -\int_{- v_{\rm F} \Lambda }^{0} \frac{d \nu}{\pi}
{\rm Im\,} 
\left[
	\left(
		G^{\rm R,0}_{\rm ret}
	\right)^2
	\Sigma^{\rm R}_{\rm ret}
\right].
\end{eqnarray}
Substituting the expressions for
$G^{\rm R,0}_{\rm ret}$
and
$\Sigma^{\rm R}_{\rm ret}$
it is easy to show that
\begin{eqnarray} 
\left(
	G^{\rm R,0}_{\rm ret}
\right)^2
\Sigma^{\rm R}_{\rm ret}
=
\frac{(g'k)^2}{8 \pi^2 v_{\rm F}^2 }
\frac{\partial} {\partial \nu}
\left[
\ln
\left(
	\frac{
		\nu - \varepsilon_k^{\rm R} + i0
	     }
	     {
		v_{\rm F} \Lambda
	     }
\right)
\right]^2
+ \ldots,
\end{eqnarray}
where, as above, the ellipsis stands for the less-singular contributions to
$\Sigma_{\rm ret}^{\rm R}$.
With the help of this formula the integral in 
Eq.~(\ref{occup_corr})
can be trivially evaluated
\begin{eqnarray}
%%%%%%%%%%%%%%%%%%%%%%%%%%%%%%%%%%%%%%%%%%%%%%%%%%
\label{delta_n}
%%%%%%%%%%%%%%%%%%%%%%%%%%%%%%%%%%%%%%%%%%%%%%%%%% 
\delta n^{\rm R}_k
\approx
\frac{(g'k)^2}{4 \pi^2 v_{\rm F}^2 }
\theta(\varepsilon_k^{\rm R})
\ln
\left(
	\frac{
		\varepsilon_k^{\rm R}
	     }
	     {
		v_{\rm F} \Lambda
	     }
\right)
+ \ldots,
\end{eqnarray}
which is finite and small for any
$|k| < \Lambda$,
provided that $g'$ is small.

\textit{Quasi-Fermi liquid.}--
%\label{qfl}
The calculations presented above prove that the quasi-Fermi liquid of 1D
spinless fermions constitute a distinct state of matter. Indeed, it is not
a Tomonaga-Luttinger liquid: since 
$\delta n^{\rm R}_k$,
Eq.~(\ref{delta_n}),
is small, the quasi-Fermi liquid occupation number is discontinuous at the
Fermi energy, while the Tomonaga-Luttinger's 
$n^p_k$
is continuous. [This dissimilarity is a consequence of the fact that the
marginal interaction in the Tomonaga-Luttinger Hamiltonian induces
stronger singularity of the
self-energy diagram than the singularity of
Eq.~(\ref{Sigma}).
As a result, for the Tomonaga-Luttinger liquid the occupation number
correction diverges for small $k$.]

On the other hand, the state of matter we are dealing with is not a Fermi
liquid because it has no perturbatively defined fermionic quasiparticles.
(Heuristic non-perturbative construction of excitations for
$H_{\rm ii}$
is discussed in Supplemental Material.)
However, the system retains certain features of the Fermi liquid: as we
have mentioned in the previous paragraph, the occupation number exhibits
finite discontinuity at 
$k=0$.
This discontinuity exists even though the quasiparticles do not.

%Above we already have mentioned that Hamiltonian $H$ of
%Eq.~(\ref{H_gen})
%can be mapped on 
%$H_{\rm ii}$.
%Indeed, using the unitary transformation (proposed in
%Ref.~\onlinecite{mattias_lieb1965})
%it has been demonstrated that the two Hamiltonians are unitary equivalent
%\cite{rozhkov2005}.
%This means that the study of the quasi-Fermi liquid is important for
%understanding the generic 1D model $H$.

Let us now discuss experimental identification of the quasi-Fermi liquid.
Due to its peculiar nature, the quasi-Fermi liquid may present itself on
experiment as an ordinary Fermi liquid, unless the measurements are
done at sufficiently high energy. Indeed, formally, the quasiparticle
residue diverges for any finite $k$, however, the divergence becomes
progressively weaker as $k$ approaches the Fermi point. 

To appreciate the latter point imagine that the single-fermion spectral
function is measured, and the quasiparticle residue is extracted. For an
experimental apparatus with finite resolution width $\Omega$ the measured
value of
$\delta Z^{{\rm R}, \Omega}_k$
is never divergent
\begin{eqnarray}
|\delta Z^{\rm R, \Omega}|
=
\frac{(g'k)^2}{4\pi^2 v_{\rm F}^2} 
\ln
\left(
	\frac{
		v_{\rm F} \Lambda
	     }
	     {
		\Omega
	     }
\right) < \infty.
\end{eqnarray} 
In this expression the divergence of
$\delta Z^{\rm R}_k$,
Eq.~(\ref{dZ}),
is cut at the energy scale 
$\sim \Omega$.
Nonetheless, it is possible that
$|\delta Z^{\rm R, \Omega}| > 1$,
provided that $k$ is not too small:
$k > k^\times$,
where 
$k^\times$
is equal to
\begin{eqnarray} 
%%%%%%%%%%%%%%%%%%%%%%%%%%%%%%%%%%%%%%%%%%%%%%%%%%
\label{kX}
%%%%%%%%%%%%%%%%%%%%%%%%%%%%%%%%%%%%%%%%%%%%%%%%%% 
k^\times
=
\frac{2\pi v_{\rm F}}
     {
	g' 
	\sqrt{
		\ln
			\left(
				\frac{
					v_{\rm F} \Lambda
	     			     }
				     {
					\Omega
				     }
			\right)
	     }
     }.
\end{eqnarray} 
The quantity
$k^\times$
defines the crossover scale: for momenta smaller than
$k^\times$
the behavior of the system is indistinguishable from the usual Fermi
liquid. Indeed:
%\begin{eqnarray}
$|k|
<
k^\times
\Leftrightarrow
|\delta Z^{{\rm R}, \Omega} | < 1$.
%\end{eqnarray}
Thus, the characteristic divergence of the quasiparticle residue may
be measured only for momenta $k$ in the interval:
$k^\times < |k| < \Lambda$.
If the resolution is so poor that 
$k^\times > \Lambda$,
the experimentally measured behavior of the system is indistinguishable
from the Fermi liquid for any $k$. This imposes a restriction on $\Omega$:
it has to be smaller than
%\begin{eqnarray}
$
\Omega_{\rm max} 
= v_{\rm F} \Lambda 
\exp
	\left[
		-\left(
			2 \pi v_{\rm F}/g' \Lambda
		\right)^2
	\right].
$
%\end{eqnarray}
Therefore, unless we have an apparatus with exponentially sharp
resolution, phenomenology of the quasi-Fermi liquid may be observed only if
$g'$ is not too small.
%Indeed, for poor resolution
%$\sqrt{\ln{ (v_{\rm F} \Lambda/ \Omega) }} \sim 1$.
%Since
%$k^\times < \Lambda$,
%the following must be true:
%\begin{eqnarray}
%%%%%%%%%%%%%%%%%%%%%%%%%%%%%%%%%%%%%%%%%%%%%%%%%%%
%\label{large_constant}
%%%%%%%%%%%%%%%%%%%%%%%%%%%%%%%%%%%%%%%%%%%%%%%%%%% 
%\bar g'
%=
%\frac{
%	g' \Lambda
%     }
%     {
%	2 \pi v_{\rm F}
%     }
%>
%1,
%\end{eqnarray}
%which implies that the formal expansion parameter 
%$\bar g'$
%is not small, 
However, at larger $g'$ our perturbation theory becomes less accurate. Can
the quasi-Fermi liquid survive in the non-perturbative regime? We
hypothesize that the quasi-Fermi liquid, much like the Fermi or
Tomonaga-Luttinger liquids, constitutes its own separate universality
class, and the quasi-Fermi liquid phenomenology extends beyond the
small-$g'$ region.

\textit{Cold atoms.}-- Finally, let us discuss possible implementation of
the quasi-Fermi liquid with the help of the cold fermion atoms in a trap
\cite{1d_fermi_atoms_2005}. 
To characterize the gas, instead of using full inter-atomic potential 
$V(x)$,
the interactions in such systems are modeled by an effective
delta-function-like potential with the corresponding coupling $g$. Such
formalism is equivalent to our
$H_{\rm int}$
[see
Eq.~(\ref{int})],
which also describes the contact interaction between the fermions.
Experimentally, it is possible to control the magnitude and sign of
coupling $g$. Moreover, $g$ can be nullified. When this nullification
occurs, however, the atoms will not behave as non-interacting gas. Indeed,
vanishing of 
$H_{\rm int}$
does not imply the vanishing of the irrelevant 
$H_{\rm int}'$,
which drives the system toward the quasi-Fermi liquid.

To be more specific, consider the following toy model: a 1D
fermions gas with weak interaction 
$\int dx dx' V(x-x')\rho(x)\rho(x')$.
%\begin{eqnarray} 
%\rho (x)
%= 
%\rho_{\rm L} (x)+ \rho_{\rm R} (x)
%+ 
%[
%\psi^\dag_{\rm R} (x)\psi^{\vphantom{\dagger}}_{\rm L} (x)e^{-2ik_{\rm F} x} 
%+
%{\rm h.c.}
%]
%\end{eqnarray} 
%near $x'$ in powers of 
%$(x-x')$
For such a situation the effective low-energy Hamiltonian of the form $H$,
Eq.~(\ref{H_gen}),
may be derived. The (bare) coupling constants are:
\begin{eqnarray}
g 
=
2\int V(x) 
\left[
	1 - \cos (2k_{\rm F}x)
\right]
dx,
%%%%%%%%%%%%%%%%%%%%%%%%%%%%%%%%%%%%%%%%%%%%%%%%%%
\label{v_marg}
%%%%%%%%%%%%%%%%%%%%%%%%%%%%%%%%%%%%%%%%%%%%%%%%%% 
\\
g' 
=
- \int x V(x) \sin (2k_{\rm F}x)
dx.
%%%%%%%%%%%%%%%%%%%%%%%%%%%%%%%%%%%%%%%%%%%%%%%%%%%
\label{v_irr}
%%%%%%%%%%%%%%%%%%%%%%%%%%%%%%%%%%%%%%%%%%%%%%%%%%% 
\end{eqnarray} 
Usually, it is enough to retain $g$, and $g'$ is discarded due to its
irrelevance.

Imagine now that we adjust $V$ to cancel $g$. [Strictly speaking, we must
eradicate the renormalized coupling 
$g^{\rm eff} = g + {\cal O}((g')^2)$;
however, when $V$ is small, the corrections to the bare coupling are
insignificant.] In a generic situation $g'$ remains finite even when 
$g = 0$.
Of course, in this case $g'$ cannot be neglected,
and
$H_{\rm ii}$
[see
Eq.~(\ref{no_marg})]
is realized. The aim of this discussion is to demonstrate that upon
destruction of the marginal interaction one does not arrive at the free
fermion theory. Rather, the new effective theory has the irrelevant
interaction term, and our system becomes the quasi-Fermi liquid. 

%%Unfortunately,
%%Eqs.~(\ref{v_marg},\ref{v_irr})
%%are not applicable when internal or transverse degrees of freedom
%%contribute substantially to the atomic scattering
%%\cite{olshanii1998,granger_blume2004,cold_atoms_rmp2008}.
%%There, the interaction between the fermions cannot be described by a single
%%potential energy $V$, and a more complicated formalism is necessary.
%%However, even in such a case,
%%Eq.~(\ref{H_gen}), 
%%remains valid as an effective low-energy Hamiltonian: since all terms of
%%Eq.~(\ref{H_gen})
%%are compatible with the symmetry of the microscopic system, then,
%%barring a coincidence,
%%$v_{\rm F}$,
%%$v'_{\rm F}$,
%%$g$, $g'$, and other even more irrelevant terms are non-zero. [Yet, values
%%of the coupling constants are not necessary given by
%%Eqs.~(\ref{v_marg},\ref{v_irr}).]
%%Usually, only the marginal terms are kept, while the irrelevant ones are
%%purged. The resultant truncated Hamiltonian is given by
%%Eq.~(\ref{H_TL}).
%
%%Experimentally, the interaction constant $g$ for the atoms in an optical
%%trap can be varied in a broad range, and even pass through zero
%%\cite{cold_atoms_rmp2008}.
%%When constant $g$ vanishes, the remaining interaction terms are all
%%irrelevant: unless some special symmetry or fine-tuning is present, the
%%absence of the marginal interaction does not imply that the irrelevant
%%terms are cancelled too. Under such circumstances the most general
%%effective Hamiltonian of the system is given by
%%Eq.~(\ref{no_marg}).

\textit{Conclusions.}--
%%%%%%%%%%%%%%%%%%%%%%%%%%%%%%%%%%%%%%%%%%%%%%%%%%
%\label{conclusions}
%%%%%%%%%%%%%%%%%%%%%%%%%%%%%%%%%%%%%%%%%%%%%%%%%% 
To conclude, we have shown that the system of 1D spinless fermions with the
irrelevant interaction is neither a Fermi liquid, nor it is a
Tomonaga-Luttinger liquid. Instead, our system constitutes a distinct state
of matter, which we propose to call the quasi-Fermi liquid. The generic
Tomonaga-Luttinger Hamiltonian with non-linear dispersion is known to be
unitary equivalent to the Hamiltonian of such quasi-Fermi liquid. In
addition, we speculated that the quasi-Fermi liquid may be realized using
the cold atoms in 1D trap. 
%We also discussed possible difficulties which
%may hinder the experimental identification of the phase.

\textit{Acknowledgements.}--
This work was supported in part by RFBR grants 
No.~11-02-00708,
12-02-00339,
12-02-92100.
Discussions with M.~Olshanii, L.~Glazman, and F.~Essler are gratefully
acknowledged.

%\appendix

%\textit{Title of App}
%\label{energy_of_loop}
%%Text of the Appendix goes here

%%%%%%%%%%%%%%%%%%%%%%%%%%%%%%%%%%%%%%%%%%%%%%%%%%
% Old-style bibliography
%%%%%%%%%%%%%%%%%%%%%%%%%%%%%%%%%%%%%%%%%%%%%%%%%% 

%\begin{thebibliography}{99}
%
%\bibitem {pust} M. Pustylnik, {\sl Phys. Rev. Lett.}
%
%\bibitem{epjb} A.V. Rozhkov, {\sl Europ. Phys. J.} {\bf B}
%
%\bibitem{teber} Sofian Teber, preprint {\sl cond-mat/?} (unpublished)
%
%
%\end{thebibliography}

%%%%%%%%%%%%%%%%%%%%%%%%%%%%%%%%%%%%%%%%%%%%%%%%%%
%   Bibtex input
%%%%%%%%%%%%%%%%%%%%%%%%%%%%%%%%%%%%%%%%%%%%%%%%%% 
\bibliographystyle{apsrev_no_issn_url.bst}
% Name of the bib file goes below:
\bibliography{irrel}

\begin{thebibliography}{43}
\bibitem[{exp()}]{explanation_mapping}
\bibinfo{note}{Note that, despite the fact that Hamiltonians $H$ and $H_{\rm
  ii}$ are unitary equivalent, they belong to different universality classes.
  As we know, the universality class is defined by asymptotic behavior of
  different propagators. If we were to calculate a TL propagator $G = \langle
  O(\tau) \exp (-\tau H) O^\dag \rangle_{\rm TL}$, we can use the unitary
  equivalence of the Hamiltonians $H = U H_{\rm ii} U^{-1}$ to express $G$ as
  follows: $G = \langle [U^{-1}O(\tau)U] \exp (-\tau H_{\rm ii}) [U^{-1}O
  U]^\dag \rangle_{\rm ii}$. This propagator, however, is not equal to the
  propagator $G_{\rm ii} = \langle O(\tau) \exp (-\tau H_{\rm ii}) O^\dag
  \rangle_{\rm ii}$, and may have different asymptotic properties.}
\end{thebibliography}

\newpage

\onecolumngrid

%\appendix

\section*{
\large
Supplemental Material for 
``One-Dimensional Fermions with neither Luttinger-Liquid nor Fermi-Liquid
Behavior"
}

\setcounter{page}{1}

%\author{A.V. Rozhkov}
%
%\affiliation{
%Institute for Theoretical and Applied Electrodynamics, Moscow,
%125412, Russia
%}
%
%\affiliation{
%Moscow Institute of Physics and Technology, Dolgoprudny, Moscow Region,
%141700, Russia
%}
%
%
%\date{\today}
%
%%\maketitle
%
%\begin{abstract}
\section*{ABSTRACT}
\begin{changemargin}{2.0cm}{2.0cm} 
{
\small
To go beyond the limitations of the second-order perturbation theory, in
this Supplementary we discuss our model using non-perturbative three-band
Hamiltonian approach. We demonstrate that the hole-like excitations have
zero overlap with hole excitations of the non-interacting system. This
means that the quasi-Fermi liquid phenomenology survives. On the other
hand, if the kinetic energy has the non-linear dispersion term
($v_{\rm F}' \ne 0$),
the particle-like excitations acquire finite lifetime due to the Cherenkov
emission process. This formally makes the quasiparticle residue $Z$ finite.
However, since in our system the Cherenkov scattering is extremely weak,
$Z$ remains small. Thus, we expect that at not too low interaction the
quasi-Fermi liquid features may be observed for a particle-like excitation,
provided that the excitation momentum is not too small.
}
\end{changemargin}
\

\

%\end{abstract}

%\maketitle

\twocolumngrid

\section{Introduction}

In the main paper we demonstrated that the second-order correction 
$\delta Z$
to the quasiparticle residue $Z$ diverges in our system. Using this
observation we argued that for the model under consideration the Laudau
theory of Fermi liquid is inapplicable. However, one might object,
hypothesizing that after an infinite-order resummation a Fermi liquid is
recovered. To address this issue we will employ here a non-perturbative
approach. It is based on a heuristic mapping of the original Hamiltonian to
the so-called three-band
Hamiltonian~\cite{three-band}.
The latter Hamiltonian can be solved exactly. We will see that for a
hole-like excitation, a quasiparticle remains poorly defined even when the
treatment is extended beyond the second-order perturbation theory. 

As for a particle-like excitations, when 
$v_{\rm F}' \ne 0$,
the Cherenkov emission of low-lying fermion-hole pairs generates finite
lifetime for such excitations. Finite lifetime caps the divergence of
$\delta Z$,
restoring the validity of the perturbation theory. That is, formally, the
quasi-Fermi liquid survives for holes, but not for particles. However, we
will demonstrate that at not too small interaction and momentum, the
suppression of the quasiparticle residue for the particle-like excitations
is very strong. Thus, we expect to observe the quasi-Fermi liquid
phenomenology for both types of excitations.

\section{Three-band Hamiltonian for a hole-like excitation}

We begin our discussion with a case of a hole-like excitation. It was
argued in
Ref.~\onlinecite{three-band}.
that dynamics of a single hole in a one-dimensional system may be described
in terms of the effective three-band Hamiltonian (see figure):
\begin{eqnarray}
%%%%%%%%%%%%%%%%%%%%%%%%%%%%%%%%%%%%%%%%%%%%%%%%%%
\label{3B}
%%%%%%%%%%%%%%%%%%%%%%%%%%%%%%%%%%%%%%%%%%%%%%%%%% 
H_{\rm 3B}=H_{\rm kin} + H_{\rm int},
\\
H_{\rm kin}
=
i v_{\rm F} \int dx
\left(
        \colon
        \psi^\dagger_{\rm L}
        \nabla\psi^{\vphantom{\dagger}}_{\rm L}
        \colon
        - 
        \colon
        \psi^\dagger_{\rm R}
        \nabla\psi^{\vphantom{\dagger}}_{\rm R}
        \colon
\right)
\\
\nonumber 
+
\int dx
\colon\!
\psi^\dag_{\rm h}
\left(
	\omega_{\rm h} - i v_k \nabla 
\right)
\psi^{\vphantom{\dagger}}_{\rm h} \colon,
\\
H_{\rm int}
=
- \tilde g_{\rm hL} \int dx \rho_{\rm L}
\colon \psi^\dag_{\rm h} \psi^{\vphantom{\dagger}}_{\rm h} \colon.
\end{eqnarray} 
Loosely speaking, this Hamiltonian describes a ``high-energy" hole 
$\psi_{\rm h}$
with momentum
$p\approx k$,
which interacts with two bands of low-lying fermionic degrees of freedom
$\psi_{\rm R}$ and $\psi_{\rm L}$.

More technically, here
$\psi_{\rm h}$
corresponds to a right-moving hole with momentum $p$ confined within ``the
high-energy band":
\begin{eqnarray}
%%%%%%%%%%%%%%%%%%%%%%%%%%%%%%%%%%%%%%%%%%%%%%%%%%
\label{high-energy}
%%%%%%%%%%%%%%%%%%%%%%%%%%%%%%%%%%%%%%%%%%%%%%%%%% 
|p-k| < P, 
\quad
0 < -k \ll \Lambda.
\end{eqnarray}
The bandwidth is $2P$, where the effective cutoff is chosen to satisfy
$0 < P < |k|$.
The bare energy of the hole is
\begin{eqnarray}
\omega_{\rm h} 
=
|\varepsilon^{\rm R}_k| = |v_{\rm F} k + v_{\rm F}'k^2|,
\end{eqnarray}
its bare velocity is equal to
\begin{eqnarray}
v_k = v_{\rm F} + 2v_{\rm F}' k < v_{\rm F}.
\end{eqnarray}
Two ``low-energy" fields,
$\psi_{\rm R}$
and
$\psi_{\rm L}$,
have their momenta bound according to the inequality (see also figure):
\begin{eqnarray}
%%%%%%%%%%%%%%%%%%%%%%%%%%%%%%%%%%%%%%%%%%%%%%%%%%
\label{low-energy}
%%%%%%%%%%%%%%%%%%%%%%%%%%%%%%%%%%%%%%%%%%%%%%%%%% 
|p| < P.
\end{eqnarray} 

The fermion states outside the bands defined by
Eqs.~(\ref{high-energy}) 
and~(\ref{low-energy})
are assumed to be either almost empty, or almost completely occupied, thus,
they are ``integrated out".

\begin{figure}\centering
\leavevmode
\epsfysize=8.5cm
\epsfbox{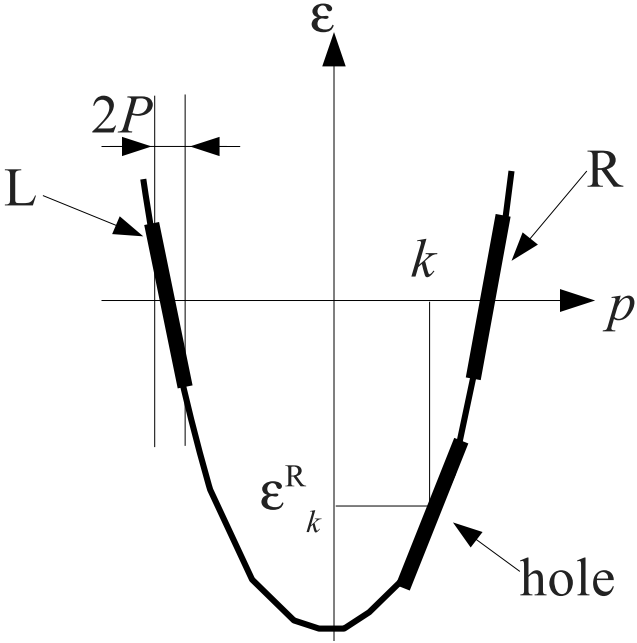}
\caption{
Kinetic energy of the three-band model. The phase space of the original
model is significantly truncated when formulating the three-band model. The
bands (`L', `R', and `hole') are shown by thick lines. Only the fermion
states within these bands are taken into account by the three-band
effective Hamiltonian. All other states are ``integrated out". The
dispersion within the bands is linearized. The ``low-energy" left-moving
(`L') and right-moving (`R') bands are centered around Fermi energy
($\varepsilon^{\rm F} = 0$
in our situation), the ``high-energy" band in which the hole is located is
centered around
$\omega_{\rm h} = \varepsilon_k^{\rm R}$.
The width of these bands is $2P$, where
$P<|k|$
serves as a cutoff of the new effective Hamiltonian.
}
\end{figure}

We assume that, due to its irrelevance, the interaction between the
low-lying fermions may be neglected (in 
Ref.~\onlinecite{three-band}
the authors studied a model with the marginal interaction; thus, they had
to retain the interaction between 
$\psi_{\rm R}$
and
$\psi_{\rm L}$).
At the same time, effective interaction between the hole and the low-lying
band is finite for finite $k$. It is characterized by the coupling constant 
\begin{eqnarray}
%%%%%%%%%%%%%%%%%%%%%%%%%%%%%%%%%%%%%%%%%%%%%%%%%%
\label{cc}
%%%%%%%%%%%%%%%%%%%%%%%%%%%%%%%%%%%%%%%%%%%%%%%%%% 
\tilde g_{\rm hL} = 2 g' k + {\cal O}(g^2).
\end{eqnarray} 
Note that due to irrelevance of the interaction, the coupling constant
vanishes when
$k \rightarrow 0$.
Yet, for any finite $k$ it remains finite. This fact is of cardinal
importance for us: scattering of the low-lying excitations by the hole
induces the orthogonality catastrophe. That is, the state of the
non-interacting system
($\tilde g_{\rm hL} = 0$)
containing one hole with momentum $k$ has zero overlap with the state of
the interacting system 
($\tilde g_{\rm hL} \ne 0$)
in which a quasi-hole with momentum $k$ is created:
\begin{eqnarray}
%%%%%%%%%%%%%%%%%%%%%%%%%%%%%%%%%%%%%%%%%%%%%%%%%%
\label{orth}
%%%%%%%%%%%%%%%%%%%%%%%%%%%%%%%%%%%%%%%%%%%%%%%%%% 
\langle k, \tilde g_{\rm hL} = 0 | k, \tilde g_{\rm hL} \ne 0 \rangle
= 0.
\end{eqnarray} 
This overlap is related to the quasiparticle residue:
\begin{eqnarray}
Z^{\rm R}_k 
=
|\langle k, \tilde g_{\rm hL} = 0 | k, \tilde g_{\rm hL} \ne 0 \rangle|^2.
%%%%%%%%%%%%%%%%%%%%%%%%%%%%%%%%%%%%%%%%%%%%%%%%%%
\label{z_vs_overlap}
%%%%%%%%%%%%%%%%%%%%%%%%%%%%%%%%%%%%%%%%%%%%%%%%%%
\end{eqnarray}
Therefore, the residue vanishes. Clearly, if 
$Z_k^{\rm R} = 0$,
the perturbation theory fails, which we have demonstrated in the main
paper. The nullification of $Z$ also disagrees with the basic assumption of
the Landau theory of a Fermi liquid.

To prove 
Eq.~(\ref{orth}) 
it is convenient to bosonize the low-lying degrees of freedom of the
three-band Hamiltonian:
\begin{eqnarray}
H_{\rm kin}
=
\frac{v_{\rm F}}{2}
\int dx
\left[
        \colon\!
        (\nabla\varphi_{+})^2
        \colon
        + 
        \colon\!
        (\nabla\varphi_{-})^2
        \colon
\right]
\\
\nonumber 
+
\int dx
\psi^\dag_{\rm h}
\left(
	\omega_{\rm h} - i v_k \nabla 
\right)
\psi^{\vphantom{\dagger}}_{\rm h},
\\
H_{\rm int}
=
-
\frac{\tilde g_{\rm hL}}{\sqrt{2\pi}}
\int dx\,
\colon\! \psi^\dag_{\rm h} \psi^{\vphantom{\dagger}}_{\rm h} \colon\!
(\nabla \varphi_{+}),
\end{eqnarray} 
where bosonic fields
$\varphi_\pm$
are related to the chiral densities:
\begin{eqnarray}
\rho_{\rm R} = - \frac{1}{\sqrt{2\pi}} \nabla \varphi_{-},
\quad
\rho_{\rm L} = \frac{1}{\sqrt{2\pi}} \nabla \varphi_{+}.
\end{eqnarray}
Using the commutation relations
\begin{eqnarray} 
\left[
	\varphi_\pm (x);
	\varphi_\pm (y)
\right]
=
\mp \frac{i}{2} {\rm sgn} (x-y),
\\
\left[
	\varphi_+ (x);
	\varphi_- (y)
\right]
=0,
\end{eqnarray} 
we can prove that for the operator $W$ defined as
\begin{eqnarray} 
W = \int dx 
\colon\! \psi^\dag_{\rm h} \psi^{\vphantom{\dagger}}_{\rm h} \colon\!
\varphi_+,
\end{eqnarray}
the following relation is valid
\begin{eqnarray}
\left[
	H_{\rm kin}; W
\right]
=
i \frac{ \sqrt{2\pi} (v_{\rm F} + v_k) }{ g_{\rm hL} } H_{\rm int}.
\end{eqnarray}
Consequently, the unitary transformation
$U_\theta={\it e}^{i \theta W}$
diagonalizes 
$H_{\rm 3B}$:
\begin{eqnarray}
{\overline H}
=
U_\theta^{\vphantom{\dagger}} H_{\rm 3B} U_\theta^\dag 
=
H_{\rm kin} + \ldots ,
\\
{\rm provided\ that\ }
\theta 
=
-\frac{ g_{\rm hL} }{ \sqrt{2\pi} (v_{\rm F} + v_k) }.
\end{eqnarray}
The ellipsis stands for correction to the bare energy of the
hole 
$\omega_{\rm h}$,
which is introduced by 
$U_\theta^{\vphantom{\dagger}} H_{\rm int} U_\theta^\dag$.

In the ground state of non-interacting Hamiltonian 
$\overline H$,
which we denote by
$\left| 0, \tilde g_{\rm hL} = 0 \right>$,
there is no hole, and all bosonic modes are in their ground states. We are
more interested, however, in the state where a hole with momentum $k$ is
present:
\begin{eqnarray} 
\left| k, \tilde g_{\rm hL} = 0 \right>
=
\int
\frac{dx}{\sqrt{L}}
\psi^\dag_{\rm h} (x) \left| 0, \tilde g_{\rm hL} = 0 \right>.
\end{eqnarray}
Of course,
$
\left| k, \tilde g_{\rm hL} = 0 \right>
$ 
is an eigenstate of 
$\overline H$.
The eigenstate of the three-band Hamiltonian 
$H_{\rm 3B}$
with a single hole and momentum $k$ is
\begin{eqnarray}
%%%%%%%%%%%%%%%%%%%%%%%%%%%%%%%%%%%%%%%%%%%%%%%%%%
\label{excited_st}
%%%%%%%%%%%%%%%%%%%%%%%%%%%%%%%%%%%%%%%%%%%%%%%%%% 
\left| k, \tilde g_{\rm hL} \ne 0 \right>
=
U^\dag_\theta
\left| k, \tilde g_{\rm hL}=0 \right>.
\end{eqnarray}

Using 
Eq.~(\ref{excited_st})
we can express the overlap from
Eq.~(\ref{orth}) 
as:
\begin{eqnarray}
\langle k, \tilde g_{\rm hL} = 0 | k, \tilde g_{\rm hL} \ne 0 \rangle
=
\\
\nonumber 
\int \frac{dx dx'}{L} 
\langle 0, \tilde g_{\rm hL} = 0 |
\psi_{\rm h}^{\vphantom{\dagger}} (x)
U^\dag_\theta
\psi_{\rm h}^\dag (x')
| 0, \tilde g_{\rm hL} = 0 \rangle.
\end{eqnarray} 
Since
$
\psi_{\rm h}^\dag \psi^{\vphantom{\dagger}}_{\rm h}
| 0, \tilde g_{\rm hL} = 0 \rangle
=0,
$
thus, this state is invariant under the transformation
$U_\theta$ for any $\theta$:
\begin{eqnarray}
U_\theta \left| 0, \tilde g_{\rm hL} = 0 \right>
=
\left| 0, \tilde g_{\rm hL} = 0 \right>.
\end{eqnarray}
This identity and the following expression for the transformed field
\begin{eqnarray}
U_\theta^\dag \psi^\dag_{\rm h} U_\theta^{\vphantom{\dagger}} 
=
\exp \left( 
		- i \theta \varphi_+
	\right)
\psi_{\rm h},
\end{eqnarray}
allow us to write
\begin{eqnarray}
\langle k, \tilde g_{\rm hL} = 0 | k, \tilde g_{\rm hL} \ne 0 \rangle
=
\\
\nonumber 
\int \frac{dx dx'}{L} 
\langle
	\psi_{\rm h}^{\vphantom{\dagger}} (x)
	\psi_{\rm h}^\dag (x')
\rangle
\langle
	\exp ( - i \theta \varphi_+)
\rangle,
\end{eqnarray} 
where
$\langle \ldots \rangle$
stands for expectation value with respect to the non-interacting ground state
$
| 0, \tilde g_{\rm hL} = 0 \rangle
$.
Since in the non-interacting system the low-lying bosons and the hole are
decoupled, the expectation value decomposes into a product of two matrix
elements, one is for the bosonic degrees of freedom, another is for the
hole. The fermionic matrix element can be evaluated quite
straightforwardly:
\begin{eqnarray} 
\langle 
	\psi_{\rm h}^{\vphantom{\dagger}} (x)
	\psi_{\rm h}^\dag (x')
\rangle
=
\delta (x-x'),
\quad
\Rightarrow
\\
\nonumber 
L^{-1} \int dx dx'
\langle 
	\psi_{\rm h}^{\vphantom{\dagger}} (x)
	\psi_{\rm h}^\dag (x')
\rangle
=
1.
\end{eqnarray} 
For the bosonic matrix element we derive:
\begin{eqnarray}
%%%%%%%%%%%%%%%%%%%%%%%%%%%%%%%%%%%%%%%%%%%%%%%%%%
\label{zero1}
%%%%%%%%%%%%%%%%%%%%%%%%%%%%%%%%%%%%%%%%%%%%%%%%%% 
\langle
	\exp ( i \theta \varphi_+)
\rangle
=
\exp\left( 
	- \frac{\theta^2}{2} \langle \varphi_+^2 \rangle 
    \right),
\\
%%%%%%%%%%%%%%%%%%%%%%%%%%%%%%%%%%%%%%%%%%%%%%%%%%
\label{zero2}
%%%%%%%%%%%%%%%%%%%%%%%%%%%%%%%%%%%%%%%%%%%%%%%%%%
\langle \varphi_+^2 \rangle 
=
\frac{1}{2\pi} {\rm ln} (P L) \rightarrow \infty,
\end{eqnarray}
when
$L \rightarrow \infty$.
Therefore,
$
\langle
	\exp ( i \theta \varphi_+)
\rangle
$
vanishes in the thermodynamic limit, and the orthogonality catastrophe
occurs.

Equations~(\ref{zero1}) and (\ref{zero2})
suggest the following interpretation of the orthogonality catastrophe: as
a result of scattering off the hole, divergent amount of ``soft"
fermion-hole pairs are excited, which leads to the nullification of the
overlap,
Eq.~(\ref{orth}),
and the quasiparticle residue,
Eq.~(\ref{z_vs_overlap}).
This renders the familiar Fermi liquid theory inapplicable.

Why irrelevant interaction has such dramatic effect on the excited state,
but not on the ground state? The irrelevant interaction in the energy
domain disappears as the Fermi energy is approached. As a result, its
effect on the ground state is amenable to the perturbation theory approach.
At the same time, the irrelevant interaction is able to generate non-zero
coupling [see
Eq.~(\ref{cc})]
for any finite-$k$ hole excitation. Due to the irrelevance of 
$H_{\rm int}$,
the coupling constant 
$\tilde g_{\rm hL}$
vanishes when
$k \rightarrow 0$,
but it remains finite for any finite $k$. This coupling is the cause of the
orthogonality catastrophe we described above.

\section{Particle-like excitation}

Thus far, we discussed the hole excitations. Let us now address the case of
particle-like excitation. Superficially, one expects that the same
orthogonality catastrophe would occur for the particle excitations as well.
This, however, is correct only when 
$v_{\rm F}'=0$.
Otherwise, since the group velocity of a particle excitation is higher than
the Fermi velocity, it acquires finite lifetime due to Cherenkov emission
of particle-hole pairs
\cite{fermi-lutt}.

The Cherenkov emission is a very weak process in our system: to satisfy
momentum and energy conservation laws, two pairs (one right-moving and one
left-moving) have to be emitted. The corresponding scattering rate in a
model with marginal interaction
$g \rho_{\rm L} \rho_{\rm R}$
has been evaluated in 
Ref.~\onlinecite{fermi-lutt}
[see Eq.~(10) in this reference]. It is proportional to the fourth power of
the interaction constant and eighth power of $k$:
$
\omega^{\rm Ch}_k
\propto 
(gg_{\rm RR}'')^2 k^{8},
$
where
$g_{\rm RR}''k^2$
characterizes the same-chirality coupling. Such a high power of $k$ is a
consequence of the fact that two particle-hole pairs must be emitted.

Since in our system the interaction between fermions of opposing
chiralities is irrelevant, we expect that
$\omega^{\rm Ch}_k$
demonstrates even faster decay. The dimensional analysis suggests that
\begin{eqnarray}
%%%%%%%%%%%%%%%%%%%%%%%%%%%%%%%%%%%%%%%%%%%%%%%%%%
\label{cherenkov}
%%%%%%%%%%%%%%%%%%%%%%%%%%%%%%%%%%%%%%%%%%%%%%%%%% 
\omega^{\rm Ch}_k
\propto 
(g'g_{\rm RR}'')^2 k^{10}.
\end{eqnarray} 
The finite scattering rate implies that the self-energy on the mass
surface acquires finite imaginary part
$\propto \omega^{\rm Ch}_k$.
It caps the divergence of the quasiparticle residue correction
$\delta Z$.
Thus, formally, for a particle-like excitation the Fermi liquid behavior
is restored, and the orthogonality catastrophe is avoided. However, due to
extreme weakness of the Cherenkov emission, the restoration of the Fermi
liquid becomes apparent only in a very narrow region near the mass surface:
\begin{eqnarray}
|\nu - \varepsilon_k^p | 
\ll
\omega^{\rm Ch}_k.
\end{eqnarray} 
Outside of this area, the quasi-Fermi liquid physics can be observed.

To be more qualitative, let us consider the correction to the quasiparticle
residue in the situation 
$
\omega^{\rm Ch}_k \ne 0
$.
The correction becomes finite:
\begin{eqnarray}
%%%%%%%%%%%%%%%%%%%%%%%%%%%%%%%%%%%%%%%%%%%%%%%%%%
\label{Z_Cherenkov}
%%%%%%%%%%%%%%%%%%%%%%%%%%%%%%%%%%%%%%%%%%%%%%%%%% 
\delta Z^{\rm R}_k
&=&
\frac{(g'k)^2}{4\pi^2 v_{\rm F}^2} 
\ln
\left(
        \frac{
                v_{\rm F} \Lambda
             }
             {
                \omega^{\rm Ch}_k
             }
\right)
\\
\nonumber 
&=&
\frac{(g'k)^2}{4\pi^2 v_{\rm F}^2} 
\times
	10 \ln \left( \frac{\Lambda}{k} \right)
+ \ldots
%\left(
%	\ln g' + \ln g_{\rm RR}'' + 10 \ln k
%\right).
\end{eqnarray} 
Neglecting weak logarithmic dependence, we write
\begin{eqnarray}
%%%%%%%%%%%%%%%%%%%%%%%%%%%%%%%%%%%%%%%%%%%%%%%%%%
\label{Z_Cherenkov_est}
%%%%%%%%%%%%%%%%%%%%%%%%%%%%%%%%%%%%%%%%%%%%%%%%%% 
|\delta Z^{\rm R}_k|
>
10\, \delta Z^{\rm \, naive}_k,
\ {\rm where\ }
\delta Z^{\rm naive}_k
=
\frac{(g'k)^2}{4\pi^2 v_{\rm F}^2}.
\end{eqnarray} 
That is, while the logarithmic divergence is absent, 
$\delta Z$
experiences strong renormalization (one order of magnitude, approximately)
as compared to its ``naive" estimate
$
\delta Z^{\rm naive}.
$
Thus, at not too small $g'$ and $k$ (that is, when
$\delta Z^{\rm naive}_k \lesssim 1$)
we can expect significant suppression of the quasiparticle residue for the
particle-like excitations. In such a regime, phenomenology of the
quasi-Fermi liquid may be observed experimentally.

\end{document}